\documentclass[showpacs,twocolumn,amsmath,amssymb,fixfloats]{revtex4}
\usepackage[english]{babel}
\usepackage{amsmath}
\usepackage{amssymb}
\usepackage{graphicx}

\begin{document}

\title{Bistable Mott-insulator to superfluid phase transition in cavity optomechanics}

\author{W. Chen$^{1}$}
\author{K. Zhang$^{2}$}
\author{D. S. Goldbaum$^{1}$}
\author{M. Bhattacharya$^{1}$}
\author{P. Meystre$^{1}$}
\affiliation{
\begin{tabular}{c}
$^{1}$B2 Institute, Department of Physics and College of Optical Sciences, The University of Arizona, Tucson, AZ 85721, USA \\
$^{2}$State Key Laboratory of Precision Spectroscopy, Department of Physics, East China Normal University, Shanghai 200062, China
\end{tabular}
}


\begin{abstract}

We study the many-body state of ultracold bosons in a bistable optical lattice potential in an optomechanical resonator in the weak-coupling limit. New physics arises as a result of bistability and discontinuous jumps in the cavity field. Of particular interest is the situation where the optical cavity is engineered so that a single input beam can result in two radically different stable ground states for the intracavity gas: superfluid and Mott-insulator. Furthermore, the system we describe can be used as an adjustable template for investigating the coupling between cavity fields, nanomechanical systems operating in the quantum regime, and ultracold atomic gases.
\end{abstract}
\maketitle

Recent years have witnessed a remarkable convergence of interests in atomic, molecular and optical physics, condensed matter physics, and nanoscience. Specific examples include the use of ultracold atomic and molecular systems as quantum simulators of solid-state systems
\cite{Jaksch:ColdBosonicAtomsinOL, Zoller:OLreview}, the demonstration of the analog of cavity QED effects with superconducting boxes~\cite{Schoelkopf:2008xr}, and the laser cooling of nanoscale cantilevers
\cite{Metzger:CavityCoolingMicrolever}, leading to the emerging field of cavity optomechanics.

The central element of most cavity optomechanical systems consists of a Fabry-P{\'e}rot type cavity with one end-mirror vibrating about its equilibrium position under the effect of radiation pressure. These devices can exhibit optical bistability, that is, the light transmitted through the cavity can take two distinct intensity values for a given incident intensity~\cite{Dorsel:OB_MirrorConfinementInducedbyRP}.

In this letter we show that optical bistability can lead to fascinating
new effects in the dynamics of an ultracold sample of bosonic atoms trapped inside such resonators. In particular, at the simplest level of weak coupling and classical mirror motion we predict a bistable quantum phase transition between a Mott-insulator (MI) state and a superfluid (SF) state of the many-atom system. In the more general case where these approximations are removed, this system opens the way to the exploration of a completely new regime of interaction between light, ultracold atoms and quantum mechanical nanostructures.

We note at the outset that clearly, a bistable transition between a MI and a SF does not require the use of a cavity optomechanical system: any arrangement producing optical bistability would work just as well. However, it is expected that it will soon be possible to efficiently laser-cool one or more modes of vibration of moving nanoscale cantilevers or mirrors to their quantum mechanical ground state. An added advantage of the optomechanical cavity setup is its ability to serve as a diagnostic: the reflected or transmitted fraction of light driving the cavity has been shown to contain information about atomic~\cite{Wenzhou_PRA_75, Ritsch_NaturePhysics_3, Brennecke:CavityOptomechanicswithBEC} and mirror~\cite{Metzger:CavityCoolingMicrolever} dynamics. This is what makes the coupling of ultracold atoms to optomechanical systems so promising. One main purpose of this note is to demonstrate that these studies are rapidly becoming experimentally viable.

We recall that the MI to SF transition can occur when an ultracold gas of bosonic atoms is trapped by an optical lattice in the tight-binding regime~\cite{Greiner:QuantumPhaseTransitionSF2MI}.
The ground state properties of the system are largely determined by the
relative strength of the interwell tunneling energy $J$, and the intrawell pair-interaction energy $U$. When tunneling dominates, the ground state tends to be SF. In the opposite case, the ground state tends to be a MI, characterized by a fixed atom number at each site.

Consider then an ultracold gas of bosonic atoms trapped in the optical lattice provided by the standing optical wave inside an optical cavity exhibiting bistability. In the lower intensity branch the optical lattice is shallow, so that interwell tunneling dominates and the many-atom ground state is SF. In the upper branch a much deeper optical lattice suppresses tunneling, and the many-atom ground state is a MI. The state of the atomic system is therefore bistable, with a SF or a MI being formed for the same incident light field, depending on the history of the system. In the following we describe a scenario where this effect can be observed for realizable parameters in optomechanical resonators.

Our study complements recent experiments on cold atomic gases in optical cavities with fixed ends. In each system two dynamical quantities are strongly coupled, necessitating a self-consistent, and generally nonlinear, description of their time evolution. Slama {\it et al.}~\cite{Slama_Combined} studied the gain mechanisms behind superradiant Rayleigh scattering and collective atomic recoil lasing by investigating a ring cavity system. Two separate groups investigated optomechanical systems, where collective excitations of the confined gas played the role of the mechanical oscillator. Brennecke {\it et al.}~\cite{Brennecke:CavityOptomechanicswithBEC} demonstrated a coupling between a density modulated Bose-Einstein condensate (BEC) and the cavity field, where the phase space evolution was mapped onto that of a harmonically confined, cavity-coupled mechanical oscillator. Gupta {\it et al.}~\cite{Gupta:MIstatesofUltracoldAtomsinOpticalResonators} and Murch {\it et al.}~\cite{Murch:ObservationofQuantumMeasurementBackatctionwithUltracoldAtomicGas} found that cavity-field coupling to a collective center-of-mass-motion excitation of the confined gas, resulted in oscillatory displacement of the gas.

Our work has an especially close correspondence with that of Larson {\it et al.}~\cite{Larson:MIstatesofUltracoldAtomsinOpticalResonators}. Like us, they investigate a cold gas of bosonic atoms trapped by a bistable optical lattice. However, in contrast to our study, their cavity had fixed ends, and the bistability results from the strong coupling between the cavity field and the atomic gas. Accordingly, their system is modeled by a Bose-Hubbard Hamiltonian characterized by the parameters $J$,$\, U$, and chemical potential $\mu$ calculated self-consistently with the many-body atomic state. This self-consistent dependence results in a radically different ground-state phase diagram than for the Bose-Hubbard Hamiltonian describing our system. Furthermore, our system has an additional dynamical component -- the movable end-mirror, and as already mentioned it can be used to investigate couplings between these three dynamical components when one, two or all of them operate in the quantum regime. In this letter we focus on the new physics in the weak coupling regime and for classical mirror motion. Understanding this limit is an important first step in the study of the more complicated regimes that can be realized in our setup.

On the microscopic level, the lattice potential results from the coupling between the intracavity field and an atomic resonance with frequencies, $\omega$ and $\omega_a$, respectively. As already mentioned we investigate the weak-coupling limit defined by $N g_0^2/ \lvert \Delta \rvert \ll \kappa$, where $N$ is the total number of atoms, $g_0$ is the atom-field coupling strength, $\kappa$ is the cavity's natural line-width, and $\Delta = \omega-\omega_a$ is the atom-field detuning~\cite{Horak:CoherentDynamicsofBECinHighFinesseOC}. In this limit the intracavity field has no significant dependence on the intracavity atomic population. We thus explain the generation of the intracavity optical lattice potential by using the theory of an empty cavity~\cite{Meystre:Theo_RP_driven_interferometer}.

\newcommand{\lcf}{E_{\text{L}}}
\newcommand{\rcf}{E_{\text{R}}}
\newcommand{\ein}{E_{\text{in}}}
\newcommand{\iin}{I_{\text{in}}}
\newcommand{\etrans}{E_{\text{trans}}}
\newcommand{\icav}{I_{\text{cav}} \negthinspace \left( x \right)}
\newcommand{\lcfx}{E_{\text{L}} \negthinspace \left( x \right)}
\newcommand{\rcfx}{E_{\text{R}} \negthinspace \left( x \right)}
\newcommand{\ires}{I_{\text{res} }}
\newcommand{\itrans}{I_{\text{trans} }}
\newcommand{\iout}{I_{\text{out}}}
We briefly derive the necessary results from the one-dimensional equilibrium theory of the Fabry-Perot cavity shown in figure 1. The cavity consists of two mirrors, one fixed along $x=0$ and the other harmonically confined about $x=L_0$. Each mirror has complex transmission and reflection coefficients $t$ and $r$, where $\lvert r \rvert^2=0.99$ and $\lvert t \rvert^2 + \lvert r \rvert^2 = 1$. We only consider internal reflections where one may assume a $\pi$-phase shift, and thus we replace the complex $r$ defined above with $-r$, where the new $r$ is positive and real. The phase of $t$ has no bearing on our results.

\begin{figure}
\includegraphics[width=6cm]{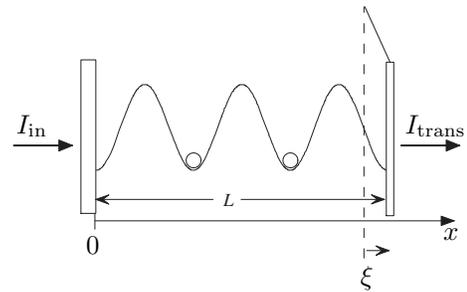}
\caption{Fabry-P{\'e}rot cavity of length $L$ with left-end mirror fixed along $x=0$ and right-end mirror oscillating about $x=L_0$, where $L=L_0+\xi$. The input- and transmitted-light intensities are labeled  $\iin$ and $\itrans$, respectively. The intracavity intensity at resonance $\left( \iin=\itrans \right)$ is represented schematically by the sine-squared wave drawn inside the cavity. In this letter, $L~\sim1$ mm long, enclosing $\sim 2000$ standing wave periods, where the magnitude of the intracavity intensity is $\sim400 \, \itrans$.  }
\label{schematic}
\end{figure}

A driving laser field $\ein$ of frequency $\omega$ is incident on, and directed normal to, the outer surface of the fixed mirror. This configuration allows a one-dimensional treatment. For a fixed cavity length, $L$, we follow the discussion of Loudon~\cite{Loudon} to determine the transmitted intensity, $\itrans$, exiting the cavity. The right-moving intracavity field at $x=0$, $\rcf $, is determined by solving $\rcf = t \ein - r \lcf$ under the equilibrium condition $\lcf = -r \exp{\negthinspace \left[ i 2 k L \right]} \, \rcf$, where $\lcf$ is the left-moving cavity field at $x=0$ and $k=\omega/c$ is the wavenumber of the light~\endnote{Note that the intracavity field relaxation time is normally much smaller than the movable mirror period for the small cavities under consideration, in which case it can be assumed to be in quasi-equilibrium. }. The resulting transmitted intensity, $\itrans=\lvert \etrans \rvert^2$, is
\begin{equation}
\itrans=\frac{\iin}{1+\frac{4\mathcal{F}^2}{\pi^2} \sin^2(k L)} \, ,
\label{aa}
\end{equation}
where $\etrans=t \rcf$, $\iin = \lvert \ein \rvert^2$, and $\mathcal{F}=\pi r / \left( 1-r^2 \right)$ is the cavity finesse.

Small mirror displacements due to the intracavity radiation pressure are given by $\xi= \eta \itrans$, where $\eta=\frac{A}{M \Omega^2 c}\frac{2 r}{\pi} \mathcal{F}$, $A$ is the cross-sectional area of the input laser beam, $M$ is the mass of the moveable mirror and $\Omega$ its oscillation frequency.
Substituting $L = L_0 + \xi$ into equation~\eqref{aa} results in a nonlinear equation for $\itrans$ which is multistable with respect to $\iin$. We concentrate on the physics near cavity resonances, where $k L = n \pi$, with $n$ a positive integer. For small displacements from resonance, the governing equation is approximately cubic in $\itrans$,
and predicts radiation pressure bistability~\cite{Dorsel:OB_MirrorConfinementInducedbyRP, Meystre:Theo_RP_driven_interferometer}.
\newcommand{\volx}{V_{\text{OL}} \negthinspace \left( x \right)}

\begin{figure}
\includegraphics[width=7cm]{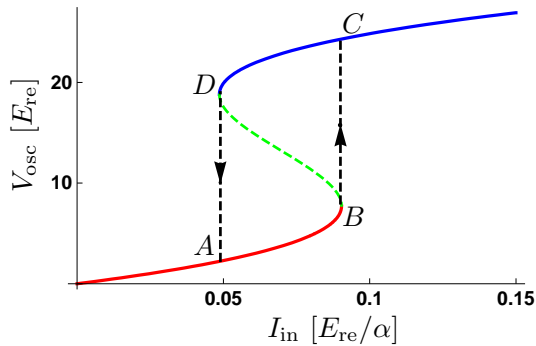}
\caption{(Color Online) Bistability of the intracavity optical lattice depth, $V_{\text{osc}}$, with respect to the input light intensity $\iin$. The bistability curve is drawn with respect to the unitless quantities $\alpha \iin/E_{\text{re}}$ and $V_{\text{osc}}/E_{\text{re}}$, where $\alpha=\volx/\icav$ and $E_{\text{re}}=\left( \hbar^2 k^2 \right)/\left( 2 m \right)$ is the recoil energy. The detuning from resonance is quantified by $\phi_0=-0.005\, \pi$, and $k \, \eta=0.1 \,  \pi \, \alpha/E_{\text{re}}$, where $\phi_0 = \text{mod}\, \pi \left[ k L_0 \right]$.
The curves $\overline{\text{AB}}$ and $\overline{\text{DC}}$ are the lower and upper branches of $V_{\text{osc}}$ in the bistable region. The dashed green line connecting D and B marks the unstable lattice depths. The dashed gray lines $\overline{\text{DA}}$ and $\overline{\text{BC}}$ mark discontinuous jumps in the lattice height. }
\label{it}
\end{figure}

It follows that the intracavity field intensity, $\icav = \lvert \rcfx + \lcfx \rvert^2$, is also bistable, and leads to a bistable optical lattice potential for the atoms~(see Figure 2),
\begin{equation}
\volx = V_{\text{osc}} \sin^2{\left[ k \left( L-x \right)\right]} + V_{\text{L}} \, ,
\label{dd}
\end{equation}
where $V_{\text{osc}} = \frac{ 4 \mathcal{F}}{ \pi } \alpha \itrans$, 
$V_{\text{L}}=  \frac{\left( 1-r \right)}{\left(1+r \right)} \alpha \itrans$, $\alpha=\left( 3 \pi c^2 \Gamma \right)/\left( 2 \omega_a^3 \Delta \right)$, and $\Gamma$ is the natural linewidth of the atomic resonance.  The microscopic origin of the proportionality constant $\alpha$ is the AC-Stark shift of the single-atom ground state. (We ignore $V_{\text{L}}$ in the following since it is tiny compared to all relevant energies.) The position of the individual lattice wells is bistable as well, since a mirror displacement, $\xi$, displaces each optical lattice well by $\xi$ in the same direction. However, we consider a regime where $\xi / \left( \pi/k \right) \sim 10^{-3}$, and thus we ignore this effect.

We consider a gas of ultracold bosonic atoms trapped in the one-dimensional optical lattice potential $\volx$. In the weak-coupling limit the atomic state does not alter the cavity field. Thus the atomic state is described by the Hamiltonian
\newcommand{\adi}{\hat{a}_{i}^{\dagger} }
\newcommand{\ai}{\hat{a}_{i} }
\newcommand{\adj}{\hat{a}_{j}^{\dagger} }
\newcommand{\aj}{\hat{a}_{j} }
\newcommand{\psidx}{\hat{\psi}^{\dagger} \negthinspace \left(  x \right) }
\newcommand{\psix}{\hat{\psi} \negthinspace \left(  x \right) }
\begin{eqnarray}
\hat{H} \negthickspace = \negthickspace \int \negmedspace dx \,\psidx \negmedspace \left( \negmedspace -\frac{\hbar^2}{2 m} \frac{d^2}{dx^2} + \volx +\frac{g}{2} \hat{n} \negthinspace \left(x \right) \negmedspace \right) \negmedspace \psix  ,
\label{a}
\end{eqnarray}
where $\psidx, \psix$ are bosonic field operators, $\hat{n} \negthinspace \left(x \right)$ is the corresponding number operator, $m$ is the atomic mass and $g$ is the two-body interaction.

We are interested in the SF--MI transition, where the many-atom system is accurately described by a tight-binding approximation that results in a single-band Bose-Hubbard Hamiltonian
\newcommand{\numi}{\hat{n}_i}
\newcommand{\aione}{\hat{a}_{i+1} }
\begin{eqnarray}
\hat{H}_{BH} \negmedspace = \negmedspace -J \sum_{\langle i,j \rangle}
\adi \aj
+ \frac{U}{2} \sum_i \numi \left( \numi - 1 \right)
-\mu   \sum_i \numi \, ,
\label{b}
\end{eqnarray}
\newcommand{\wani}{w \negthinspace \left( x-x_i \right)}
where $\adi$ $\left( \ai \right)$ is the bosonic creation (annihilation) operator for site $i$, $\numi=\adi \ai$, and the subscript $\langle i,j \rangle$ denotes a sum over nearest neighbor hopping moves. The tunneling matrix element is $J$, $U$ is the pair interaction energy, and $\mu$ is the chemical potential. The parameters $J$ and $U$ are calculated by expanding the boson field operators in a basis of lowest band Wannier states, $\psix=\sum_i \ai \wani$,
and then evaluating the pertinent integrals~\cite{Jaksch:ColdBosonicAtomsinOL}.

The ground state of the many-body system described by equation~\eqref{b} is largely determined by the value of $J/U$, which depends on the intensity, wavelength and detuning from atomic resonance of the intracavity standing wave field. Of particular interest to us is the bistable regime of the optical lattice potential where the many-body ground state corresponding to the lower branch of the potential is a SF, while the ground state corresponding to the upper branch is a MI.

Figure 3 summarizes key features of the system for a cavity length $L=1$ mm and a moving end mirror of mass $M=10$ mg and oscillation frequency $\Omega=2 \pi \times \left( 25 \, \text{Hz} \right)$. The cavity is loaded with a Bose-Einstein condensate of about 1000 sodium-23 atoms. We use an input laser of wavelength $\lambda=985$ nm to generate the intracavity optical lattice potential. The optical lattice consists of about 2000 sites, however we neglect the effects of direct atom-mirror interactions by assuming that only $\sim1000$ sites near the center of the cavity are appreciably populated. This situation can be realized by using a gentle additional confining potential, and results in a lattice system with an average single site population near unity.

The solid black curve in Fig.~3 is the mean-field SF--MI phase boundary on the axes $\mu/U$ and $\log_{10}\left( 2 J/U \right)$~\cite{Fisher:BosonLocalizationandSuperfluidInsulatorTransition},~\endnote{Mean-field theory does not capture strong-correlation effects, or produce high accuracy numerical results, but is sufficient to investigate novel qualitative features of our bistable many-body system.}. For $J/U$ smaller than this boundary the ground state is the single-particle MI, otherwise the ground state is SF. This diagram is overlayed by a plot of $\log_{10}\left( 2 J/U \right)$ versus $\iin$ for the many-body system described above, the logarithmic scale reflecting the exponential dependence of tunneling on intensity. The lower and upper branches are labeled with points $\left\{ \text{A,B} \right\}$ and $\left\{ \text{D,C} \right\}$, respectively. These labeled points correspond to the lattice depths labeled in Fig.~2. For $\iin$ just above zero, the lattice potential is too shallow for the system to be described by a single-band tight-binding limit. Thus assigning a value of $J/U$ is meaningless there. However, by adiabatically increasing the intensity of the input laser, the condensate settles into a single band of the optical lattice potential. For high enough lattice intensity, the system enters the tight-binding limit.
However, at the low-intensity edge of the bistable region, labeled A in Fig.~3, the system is very near the single-band tight-binding regime. That is, at point A a treatment with Eq.~\eqref{b} is appropriate for our present purpose, but a future in-depth calculation will require including higher band effects. At point B, in contrast, the system is safely in the single-band limit, and is accurately described by Eq.~\eqref{b}. It should be noted that the semi-log bistability plot has no direct correspondence to $\mu$ in Fig.~3. We merely specify that the system is prepared so that the upper branch of the bistability region lies inside the Mott-lobe, while the lower branch corresponds to a SF ground state.

\begin{figure}
\includegraphics[width=6cm]{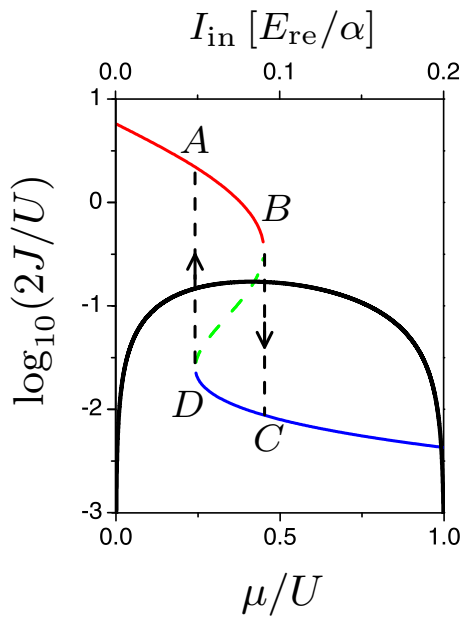}
\caption{(Color Online) Bistability of the many-body ground state. Thick black line: mean-field SF--MI phase boundary with respect to $\mu/U$ and $\log_{10}\left( 2 J/U \right)$). The ground state is the single-particle MI inside the lobe and a SF outside. The phase plot is overlayed with the intersite tunneling bistability curve, $\log_{10}\left( 2 J/U \right)$ vs. $\iin$ curve.
The ($\overline{\text{DC}}$) branch corresponds to MI ground states, while the ($\overline{\text{AB}}$) branch corresponds to SF ground states. The labeled values here correspond to the lattice depths labeled in Fig.~2. The dashed green line indicates unstable solutions, and the arrows $\overline{\text{BC}}$ and $\overline{\text{DA}}$ indicate the discontinuous jumps between different branches.  }
\end{figure}

For input intensities between the points A, D ($\iin \sim 0.86$ mW for our choice of parameters) and B,C ($\iin \sim 1.62$ mW), the system is bistable. During an initial adiabatic intensity increase, the system first resides in the lower branch, where the ground state is SF. Above point B there is only a single stable state, in the upper branch of intracavity intensity and lower branch of interwell tunneling. At that point the atoms experience a much stronger lattice confinement, with a discontinuous phase transition to a MI.

The time scale over which this transition occurs is determined by the longest of the interwell tunneling times $\tau \sim \hbar/J$ and the switching time of the intracavity field. In most cases, the intracavity field reaches a new steady-state value following an abrupt change in the incident field after a time of the order of the inverse cavity decay rate $\kappa \simeq c|t|^2/L$~\cite{PM_Bistability_1978}. However, for intensities switched from below point B to a value just above it the system undergoes a critical slowing down~\cite{PM_CriticalSlowing_1979}, with a large delay before the field switches from the lower to the upper branch. The resulting possibility to vary the switching time of the light field compared to the tunneling time provides an important tool to investigate a variety of dynamical phenomena. Most optical lattice experiments are performed using adiabatic tuning of the lattice height in an attempt to keep the system in its ground state. After sweeping through the discontinuity, though, we expect that in general the many-body state will be excited above the ground state corresponding to the optical lattice potential. The nature of this excited state and its relaxation pathways are a subject of current research. Alternatively, applying a time-dependent incident field such that the system oscillates about the discontinuity provides an additional tool to probe non-equilibrium properties, and perhaps induce coupling between MI and SF ground states.

Similar considerations hold when initially preparing a stable state in the upper branch, and then decreasing $\iin$ past point D, the optical lattice magnitude discontinuously jumps to its lower branch value at point A.

In general, the setup that we described can be used to investigate the dynamics of coupled cold atomic gases, cavity fields and nanomechanical dynamics. We considered explicitly the weak-coupling limit where the coupling between the  cavity-field and the movable mirror results in a bistable optical lattice potential for the atoms.  We have discussed how such a cavity plus cold-atom system can be engineered so that SF and MI phases are bistable ground states for the cold-atom gas. Future work will extend these considerations to the situation where the mirror motion is quantized, and discuss in detail the dynamics of the coupled system of light, ultracold atoms and quantized nanostructure both in the weak and the strong-coupling regime. With these considerations in mind an important first step is to construct an experimentally viable template, where the basic physics of each constituent system (cold gas, cavity-field, moving mirror) is well understood, and one can tune the couplings. The setup presented above is ideal for this purpose.

This work is supported in part by the US Office of Naval Research,
by the National Science Foundation, and by the US Army Research
Office.


\begin{thebibliography}{21}
\expandafter\ifx\csname natexlab\endcsname\relax\def\natexlab#1{#1}\fi
\expandafter\ifx\csname bibnamefont\endcsname\relax
  \def\bibnamefont#1{#1}\fi
\expandafter\ifx\csname bibfnamefont\endcsname\relax
  \def\bibfnamefont#1{#1}\fi
\expandafter\ifx\csname citenamefont\endcsname\relax
  \def\citenamefont#1{#1}\fi
\expandafter\ifx\csname url\endcsname\relax
  \def\url#1{\texttt{#1}}\fi
\expandafter\ifx\csname urlprefix\endcsname\relax\def\urlprefix{URL }\fi
\providecommand{\bibinfo}[2]{#2}
\providecommand{\eprint}[2][]{\url{#2}}

\bibitem{Jaksch:ColdBosonicAtomsinOL}
D. Jaksch {\it et~al.}, Phys. Rev. Lett. \textbf{81}, 3108 (1998).

\bibitem[{\citenamefont{Jaksch and Zoller}(2005)}]{Zoller:OLreview}
\bibinfo{author}{\bibfnamefont{D.}~\bibnamefont{Jaksch}} \bibnamefont{and}
  \bibinfo{author}{\bibfnamefont{P.}~\bibnamefont{Zoller}},
  \bibinfo{journal}{Ann. Phys.} \textbf{\bibinfo{volume}{315}},
  \bibinfo{pages}{52} (\bibinfo{year}{2005}).

\bibitem[{\citenamefont{Schoelkopf and Girvin}(2008)}]{Schoelkopf:2008xr}
\bibinfo{author}{\bibfnamefont{R.~J.} \bibnamefont{Schoelkopf}}
  \bibnamefont{and} \bibinfo{author}{\bibfnamefont{S.~M.}
  \bibnamefont{Girvin}}, \bibinfo{journal}{Nature}
  \textbf{\bibinfo{volume}{451}}, \bibinfo{pages}{664} (\bibinfo{year}{2008}).


\bibitem[{\citenamefont{Metzger and
  Karral}(2004)}]{Metzger:CavityCoolingMicrolever}
\bibinfo{author}{\bibfnamefont{C.~H.} \bibnamefont{Metzger}} \bibnamefont{and}
  \bibinfo{author}{\bibfnamefont{K.}~\bibnamefont{Karral}},
  \bibinfo{journal}{Nature} \textbf{\bibinfo{volume}{432}},
  \bibinfo{pages}{1002} (\bibinfo{year}{2004}).

\bibitem{Dorsel:OB_MirrorConfinementInducedbyRP}
A. Dorsel {\it et~al.}, Phys. Rev. Lett. \textbf{51} 1550 (1983).

\bibitem{Wenzhou_PRA_75}
W. Chen, D. Meiser and P. Meystre, Phys. Rev. A \textbf{75}, 023812 (2007).


\bibitem{Ritsch_NaturePhysics_3}
I. B. Mekhov, C. Maschler, and H. Ritsch, Nature Phys. \textbf{3}, 319 (2007).

\bibitem[{\citenamefont{Brennecke et~al.}(2008)\citenamefont{Brennecke, Ritter,
  Donner, and Esslinger}}]{Brennecke:CavityOptomechanicswithBEC}
\bibinfo{author}{\bibfnamefont{F.}~\bibnamefont{Brennecke}},
  \bibinfo{author}{\bibfnamefont{S.}~\bibnamefont{Ritter}},
  \bibinfo{author}{\bibfnamefont{T.}~\bibnamefont{Donner}}, \bibnamefont{and}
  \bibinfo{author}{\bibfnamefont{T.}~\bibnamefont{Esslinger}},
  \bibinfo{journal}{Science} \textbf{\bibinfo{volume}{322}},
  \bibinfo{pages}{235} (\bibinfo{year}{2008}).

\bibitem{Greiner:QuantumPhaseTransitionSF2MI}
Greiner {\it et~al.}, Nature \textbf{415}, 39 (2002).

\bibitem{Slama_Combined}
Slama {\it et~al.}, Phys. Rev. Lett. \textbf{98}, 053603 (2007); S. Slama {\it et~al.}, Phys. Rev. A \textbf{75}, 063620 (2007).

\bibitem[{\citenamefont{Gupta et~al.}(2007)\citenamefont{Gupta, Moore, Murch,
  and Stamper-Kurn}}]{Gupta:MIstatesofUltracoldAtomsinOpticalResonators}
\bibinfo{author}{\bibfnamefont{S.}~\bibnamefont{Gupta}},
  \bibinfo{author}{\bibfnamefont{K.}~\bibnamefont{Moore}},
  \bibinfo{author}{\bibfnamefont{K.}~\bibnamefont{Murch}}, \bibnamefont{and}
  \bibinfo{author}{\bibfnamefont{D.}~\bibnamefont{Stamper-Kurn}},
  \bibinfo{journal}{Phys. Rev. Lett.} \textbf{\bibinfo{volume}{99}},
  \bibinfo{pages}{213601} (\bibinfo{year}{2007}).

\bibitem[{\citenamefont{Murch et~al.}(2008)\citenamefont{Murch, Moore, Gupta,
  and
  Stamper-Kurn}}]{Murch:ObservationofQuantumMeasurementBackatctionwithUltracol%
dAtomicGas}
\bibinfo{author}{\bibfnamefont{K.}~\bibnamefont{Murch}},
  \bibinfo{author}{\bibfnamefont{K.}~\bibnamefont{Moore}},
  \bibinfo{author}{\bibfnamefont{S.}~\bibnamefont{Gupta}}, \bibnamefont{and}
  \bibinfo{author}{\bibfnamefont{M.}~\bibnamefont{Stamper-Kurn}},
  \bibinfo{journal}{Nature Physics} \textbf{\bibinfo{volume}{4}},
  \bibinfo{pages}{561} (\bibinfo{year}{2008}).

\bibitem[{\citenamefont{Larson et~al.}(2008)\citenamefont{Larson, Damski,
  Morigi, and Lewenstein}}]{Larson:MIstatesofUltracoldAtomsinOpticalResonators}
\bibinfo{author}{\bibfnamefont{J.}~\bibnamefont{Larson}},
  \bibinfo{author}{\bibfnamefont{B.}~\bibnamefont{Damski}},
  \bibinfo{author}{\bibfnamefont{G.}~\bibnamefont{Morigi}}, \bibnamefont{and}
  \bibinfo{author}{\bibfnamefont{M.}~\bibnamefont{Lewenstein}},
  \bibinfo{journal}{Phys. Rev. Lett.} \textbf{\bibinfo{volume}{100}},
  \bibinfo{pages}{050401} (\bibinfo{year}{2008}).

\bibitem[{\citenamefont{Horak et~al.}(2000)\citenamefont{Horak, Barnett, and
  Ritsch}}]{Horak:CoherentDynamicsofBECinHighFinesseOC}
\bibinfo{author}{\bibfnamefont{P.}~\bibnamefont{Horak}},
  \bibinfo{author}{\bibfnamefont{S.}~\bibnamefont{Barnett}}, \bibnamefont{and}
  \bibinfo{author}{\bibfnamefont{H.}~\bibnamefont{Ritsch}},
  \bibinfo{journal}{Phys. Rev. A} \textbf{\bibinfo{volume}{61}},
  \bibinfo{pages}{033609} (\bibinfo{year}{2000}).

\bibitem[{\citenamefont{Meystre et~al.}(1985)\citenamefont{Meystre, {E. M.
  Wright}, {J. D. McCullen}, and {E.
  Vignes}}}]{Meystre:Theo_RP_driven_interferometer}
\bibinfo{author}{\bibfnamefont{P.}~\bibnamefont{Meystre}},
  \bibinfo{author}{\bibnamefont{{E. M. Wright}}},
  \bibinfo{author}{\bibnamefont{{J. D. McCullen}}}, \bibnamefont{and}
  \bibinfo{author}{\bibnamefont{{E. Vignes}}}, \bibinfo{journal}{J. Opt. Soc.
  Am. B} \textbf{\bibinfo{volume}{2}}, \bibinfo{pages}{1830}
  (\bibinfo{year}{1985}).

\bibitem[{\citenamefont{Loudon}(2003)}]{Loudon}
\bibinfo{author}{\bibfnamefont{R.}~\bibnamefont{Loudon}},
  \emph{\bibinfo{title}{The Quantum Theory of Light}}
  (\bibinfo{publisher}{Oxford Science Publications}, \bibinfo{year}{2003}),
  \bibinfo{edition}{3rd} ed.

\bibitem[{\citenamefont{{M. P. A. Fisher} et~al.}(2000)\citenamefont{{M. P. A.
  Fisher}, {P. B. Weichman}, Grinstein, and {D. S.
  Fisher}}}]{Fisher:BosonLocalizationandSuperfluidInsulatorTransition}
\bibinfo{author}{\bibnamefont{{M. P. A. Fisher}}},
  \bibinfo{author}{\bibnamefont{{P. B. Weichman}}},
  \bibinfo{author}{\bibfnamefont{G.}~\bibnamefont{Grinstein}},
  \bibnamefont{and} \bibinfo{author}{\bibnamefont{{D. S. Fisher}}},
  \bibinfo{journal}{Phys. Rev. A} \textbf{\bibinfo{volume}{61}},
  \bibinfo{pages}{033609} (\bibinfo{year}{2000}).

\bibitem[{\citenamefont{Meystre}(1978)}]{PM_Bistability_1978}
\bibinfo{author}{\bibfnamefont{P.}~\bibnamefont{Meystre}},
  \bibinfo{journal}{Optics Commun.} \textbf{\bibinfo{volume}{27}},
  \bibinfo{pages}{147} (\bibinfo{year}{1978}).

\bibitem[{\citenamefont{Bonifacio and Meystre}(1979)}]{PM_CriticalSlowing_1979}
\bibinfo{author}{\bibfnamefont{R.}~\bibnamefont{Bonifacio}} \bibnamefont{and}
  \bibinfo{author}{\bibfnamefont{P.}~\bibnamefont{Meystre}},
  \bibinfo{journal}{Optics Commun.} \textbf{\bibinfo{volume}{29}},
  \bibinfo{pages}{131} (\bibinfo{year}{1979}).

\end{thebibliography}
\end{document}